\documentclass[aps,prd,showpacs,preprint,nofootinbib]{revtex4-1}
\usepackage{amsmath,amsthm,amssymb,verbatim,graphicx,color}

\usepackage[colorlinks]{hyperref}

\def\bea{\begin{eqnarray}}
\def\eea{\end{eqnarray}}
 \def\be{\begin{equation}}
\def\ee{\end{equation}}

\allowdisplaybreaks

\long\def\symbolfootnote[#1]#2{\begingroup%
\def\thefootnote{\fnsymbol{footnote}}\footnote[#1]{#2}\endgroup}

\newcommand{\one}{\mathbb{I}}

\newcommand{\Tr}{\mathrm{Tr}}

\newcommand{\so}{\mathrm{so}}
\newcommand{\su}{\mathrm{su}}
\newcommand{\SO}{\mathrm{SO}}
\newcommand{\SU}{\mathrm{SU}}
\newcommand{\U}{\mathrm{U}}

\newcommand{\bSigma}{\overline{\Sigma}}
\newcommand{\bsigma}{\overline{\sigma}}

\begin{document}



\centerline{\large\bf  
New ultraviolet operators in supersymmetric $\SO(10)$ GUT
}

\vskip 0.1cm

\centerline{\large\bf  
and 
consistent cosmology
}

\vskip 1.0 cm

\centerline{
\bf Piyali Banerjee\symbolfootnote[1]{
banerjee.piyali3@gmail.com
}
and
Urjit A. Yajnik\symbolfootnote[2]{
yajnik@iitb.ac.in
}
}
\medskip
\centerline{\it Department of Physics, Indian Institute of 
Technology Bombay, Mumbai 400076, India}
\bigskip
\bigskip
\begin{center}
{\large \bf Abstract}
\end{center}

We consider the minimal supersymmetric grand unified model (MSGUT)
based on the group $\SO(10)$, and study conditions leading to possible
domain wall (DW) formation.  It has been shown earlier that the supersymmetry 
preserving vacuum expectation values (vev's) get mapped to distinct but degenerate set of
vev's under action of $D$ parity, leading to formation of domain walls
as topological pseudo-defects.
The metastability of such walls can make them relatively long lived and contradict
standard cosmology. Thus we are led
to consider adding a nonrenormalisable Planck scale suppressed  operator, 
that breaks $\SO(10)$ symmetry but preserves Standard Model symmetry. For a large range of
right handed breaking scales $M_R$, this is shown to give rise to the required pressure 
difference to remove the  domain walls without conflicting with consistent big bang 
nucleosynthesis (BBN) while avoiding gravitino overproduction. However, if the walls 
persist till the onset of weak (thermal) inflation, then a low $\sim 10 - 100$ TeV $M_R$
becomes problematic.
\medskip
\medskip

PACS Numbers: 
\vfill

\section{Introduction}
All the elegant  features of both $\SO(10)$ grand unified theory (GUT) and supersymmetric models
can be incorporated in a Minimal Supersymmetric GUT (MSGUT). 
It naturally accommodates a heavy right handed neutrino in order
to fill the $\mathbf{16}$ 
representation of $\SO(10)$
with matter fermions. Thus it naturally allows  implementation of the 
seesaw mechanism \cite{Mohapatra:1980}
which explains the small non-zero neutrino masses observed 
experimentally \cite{Fukuda:2001}. 
Supersymmetric models 
with left-right symmetric gauge groups 
($\SU(3)_c \times \SU(2)_L \times \SU(2)_R \times \U(1)_{B-L}$)
naturally descend from $\SO(10)$ GUT. These models enjoy R-parity 
($R = (-1)^{3(B-L)+2S}$) \cite{Mohapatra:1986}
conservation, and so predict a stable Lightest Supersymmetric Particle 
(LSP) which is a promising dark matter 
candidate. While no signatures of low energy supersymmetry have been found at the LHC so far, 
supersymmetry continues to be an appealing mechanism for a consistent  UV completion for the 
Standard Model (SM), 
and justifies the considerations here which relate to the effects of this model in the very early universe.

An important feature of many GUT's is the occurrence of topological defects which can dynamically get formed
in the early universe~\cite{Kibble:1980,Kibble:1995}.
Although the relevant covering group $\mathrm{Spin}(10)$ is simply connected, discrete symmetry arises 
at intermediate scales in several of the sequential  breaking schemes of $\SO(10)$.  
The origin of this is the D-parity operation of the $\mathrm{Spin}(10)$
theory, which generates a discrete subgroup that becomes a $Z_2$ subgroup
of the stability group in some breaking schemes. That such breaking 
schemes can give rise to metastable domain walls, and its implications 
for cosmology were studied by Kibble, Lazarides and  Shafi~\cite{Kibble:1982}.  
A wide class of possibilities for metastable defects of various dimensions were studied
by Preskill and Vilenkin~\cite{Preskill:1993}.  The main issue is that although the parent theory has
a simply connected gauge group, the effective theory allows the formation of defects, which
cannot unwind due to the large energy scale difference between the two sequential breakings.
 In all these cases quantum tunnelling remains a possible mechanism for the destabilisation 
of such defects due to simply connected nature of the original group.

In the supersymmetric case of $\SO(10)$ MSGUT a peculiar situation arises. 
The more conventional breaking scheme proceeds via Pati-Salam (PS) 
group times D-parity ($\SU(4)_c \times \SU(2)_L \times \SU(2)_R \times D$).
This D-parity has the appealing consequence of allowing identical gauge couplings for the two 
$\SU(2)$'s allowing a chiral yet parity symmetric model of particle physics. 
The domain wall produced at the breaking of this 
$Z_2$ poses a challenge to cosmology but it could be addressed for example
by adding a parity odd singlet~\cite{Chang:1983fu} or by linking the breaking of parity to the supersymmetry 
(SUSY) breaking~\cite{Mishra:2009}.  However the scheme of breaking $\SO(10)$
directly to SM proposed by Bajc et al.~\cite{Bajc:2004}
has an unusual feature that the unique electric charge conserving vacuum expectation value (vev) 
that preserves the SM gauge group and SUSY gives an alternate 
D-parity flipped vev that also conserves charge and preserves SUSY, but 
with $\SU(2)_L$ replaced by $\SU(2)_R$ as would be manifested by fermion 
charge assignment. The domain walls then separate not gauge equivalent 
vacua but subgroups that are mirror images of each other. Further, as 
was shown in some detail in~\cite{Garg:2018}, there is  
no path that is F-flat or D-flat (in SUSY sense) that connects the two disjoint points 
of the coset space continuously. The resulting walls will be metastable 
more directly than those studied in \cite{Kibble:1982} and \cite{Preskill:1993}, since the 
defects are subject to disintegration in the course of the same phase transition 
where they are formed.  However depending on the nature of the phase transition and the rate of 
Hubble expansion, such defects begin to characterise robust local minima separated from 
the trivial vacuum by a substantial barrier which cannot be overcome once the Universe 
cools substantially below the scale of the phase transition. This class of defects was dubbed topological
pseudo-defects~\cite{Garg:2018} (TPD). Particularly, close to the Planck scale the expansion of 
the Universe would be too rapid and could compete with the time scale 
of their disappearance and lead to interesting consequences. Their 
persistence to later epochs of the Universe would be inconsistent with 
the current observed universe. Earlier works on the supersymmetric 
$\SO(10)$ GUT model studied here have not addressed this issue. 

One way to remove these domain walls 
is to introduce Planck scale suppressed non-renormalizable operators 
\cite{Rai:1994, Lew:1993}
that causes instability to domain walls. However this introduction of 
Planck suppressed operators may not work for all cases of SUSY 
and gauge symmetric models. In the Next to Minimal Supersymmetric 
Standard Model, the problem was found to persist 
\cite{Abel:1995}
in the sense that the gauge hierarchy 
problem does not get addressed if the operators required to 
remove the domain walls are permitted.
In the Supersymmetric Left-Right symmetric Models (SUSYLR) with all Higgs 
carrying gauge charges, it is possible to introduce Planck scale 
suppressed terms that are well regulated. One can then demand that 
the new operators ensure sufficient pressure across the domain walls 
that the latter disappear before BBN. 
This requirement has been discussed in detail in \cite{Mishra:2010}
in the context of R-parity conserving SUSYLR models 
\cite{Aulakh:1998}.
Similar analysis was shown to place constraints also on R-parity 
violating SUSYLR models 
\cite{Borah:2011}.

We begin by discussion of discrete set of vacua  degenerate with that signaled by 
the standard SM preserving  vevs that are used in the first stage of breaking of $\SO(10)$ \cite{Bajc:2004}.  
Applying D-parity to them gives us a new set of supersymmetry preserving SM preserving vevs. The two 
sets of vevs are disconnected in the sense that they are separated by
a potential barrier \cite{Garg:2018}. This fact leads to the danger of domain walls  formed as  topological 
pseudo-defects.

The aim of this paper is to study whether this degeneracy can be lifted using new operators in the superpotential.
Since generic dimension 3 terms have all been used up in the superpotential, it is natural to search for non-renormalisable 
ones. We next argue that adding such $\SO(10)$ preserving terms to
the superpotential will {\em not} lead to a pressure difference across any
domain wall that can arise, and thus, cannot contribute to the required 
instability. This forces us to go beyond the cherished principle of exact $SO(10)$
gauge symmetry, in particular sacrifice some of its features related to ultraviolet (UV) 
completion. The mildest such violation could be considered along the lines
of Gell-Mann \cite{GellMann:1961} and Okubo \cite{Okubo:1962} 
and add $\SO(10)$ breaking but SM preserving terms to the superpotential. In order to not touch the relevant and
marginal operators, the leading new term in the superpotential could be of mass dimension 
$4$, suppressed by one power of the cutoff scale of the theory  which we take to be the Planck scale $M_{Pl}$. 
We implement our idea by finding a coefficient 
matrix for the dimension four term drawn from the $\SO(10)$ group that
commutes with SM gauge group embedded into $\SO(10)$.  The systematically developed Pati-Salam embedding\cite{Aulakh:2005}
proves to be a good computational aid for this purpose . 
This leads to a novel ultraviolet operator that can potentially 
cause sufficient instability and remove any domain wall that may arise.

Towards consistency check for such a scenario, we demand
that the effects remain confined to UV limit approaching the Planck scale, and
naturalness of such deviation from full $SO(10)$ is ensured by the coefficients
of such terms remaining small. This naturalness test is carried out by considering the 
implications of such pressure difference to cosmology, specifically to ensuring
that the DW disappear without conflicting with standard cosmology. This is done by
testing these terms for three different scenarios
of domain wall dynamics and their eventual removal during three possible kind
of cosmological epochs, viz. radiation dominated era,
early stage of matter dominated era and late stage of matter dominated
era following a period of weak inflation. Domain wall removal during
the same cosmological periods for the cases  of left right 
symmetric supersymmetric models was earlier studied in \cite{Mishra:2010} and \cite{Borah:2011}. 
We calculate the conditions for successful domain wall removal during the three stated 
cosmological periods  in our extended MSGUT model, leading to constraints on the 
scale of right hand symmetry breaking characterised by mass scale $M_R$. Generically it 
results in upper bounds on $M_R$, in fact forcing it to remain small compared to generic GUT scale.

The paper is organised as follows. In Sec. \ref{sec:degeneratevacua} we show the existence of $D$-flipped degenerate vacua, and
the inadequacy of $SO(10)$ invariant terms in lifting this degeneracy. In sec. \ref{sec:nonrenormalisable} we propose our strategy for
overcoming the problem along the lines of octet dominance in old hadron flavour physics, and identify the required $SO(10)$ matrix.
Section \ref{sec:effectivepotential} computes the effective potential terms arising from the new nonrenormalisable operator, with details in
appendix \ref{sec:appA}.
In its subsection \ref{sec:DWremoval} we impose the required conditions on the pressure difference terms to ensure consistent cosmology
and obtain constraints on the scale of right handed symmetry breaking $M_R$. The final section \ref{sec:conclusion} contains conclusions.

\section{MSGUT superpotential and the degenerate MSSM vacua }
\label{sec:degeneratevacua}
The heavy Higgs part of the renormalisable superpotential of
$\SO(10)$ MSGUT is as follows
\cite{Bajc:2004}:
\be
W_{\mathrm{ren}} =
\frac{m}{4!} \Phi_{ijkl} \Phi_{ijkl} +
\frac{\lambda}{4!} \Phi_{ijkl}\Phi_{klmn}\Phi_{mnij} +
\frac{M}{5!} \Sigma_{ijklm} \bSigma_{ijklm} +
\frac{\eta}{4!} \Phi_{ijkl} \Sigma_{ijmno} \bSigma_{ijmno},
\ee
where $\Phi$ is the $4$-index anti-symmetric representation
$\mathbf{210}$, 
$\Sigma$ is the $5$-index self-dual anti-symmetric representation 
$\mathbf{126}$,
and $\bSigma$ is the $5$-index anti-self-dual anti-symmetric representation 
$\mathbf{\overline{126}}$
of $\SO(10)$. In the above expression, all indices range
independently from $1$ to $10$. Below, for brevity of notation, we
shall use $0$ to denote the index $10$ in our expressions.

In order to break $\SO(10)$ while preserving supersymmetry, 
we consider the following vevs. These
vevs are uniquely determined by the requirement that they be singlet
under the SM gauge group embedded into $\SO(10)$ via Pati-Salam and satisfy
electric charge conservation \cite{Bajc:2004}.
\begin{equation}
\label{eq:bajcvev}
\begin{array}{c}
\Phi_{1234} = \Phi_{1256} = \Phi_{3456} = a, \\
\Phi_{1278} = \Phi_{129\,10} = \Phi_{3478} =
\Phi_{3490} = \Phi_{5678} = \Phi_{5690} = \omega, \\
\Phi_{7890} = p,\\
\\
\Sigma_{a+1,b+3,c+5,d+7,e+9} = i^{a+b+c-d-e} \frac{\sigma}{2^{5/2}},
~~~~ a, b, c, d, e \in \{0,1\}, \\
\\
\bSigma_{a+1,b+3,c+5,d+7,e+9} = i^{-a-b-c+d+e} \frac{\bsigma}{2^{5/2}},
~~~~ a, b, c, d, e \in \{0,1\}.
\end{array}
\end{equation}
Remaining coordinates are set to zero in the vevs.

Let us consider the exchange of the $SU(2)_L$ and $SU(2)_R$ subgroups in this embedding. It amounts to the following element of $\SO(10)$ 
which achieves the desirable left-right (LR) flipping. 
\cite[Equations~28,29]{Aulakh:2005}:
\be
(L \leftrightarrow R)_{ij} =
\begin{array}{r l}
0 & \mbox{$i \neq j$} \\
-1 & \mbox{$i = j = 0$} \\
1 & \mbox{$i = j = 1, \ldots, 9$}.
\end{array}
\ee
This flipping can be effectively obtained by utilising the $D$-parity operator defined to be,
\be
D=\exp(-i\pi J_{23})\exp(i\pi J_{67})
\ee
In the present embedding, it amounts to the following element of $\SO(10)$ 
\cite[Equation~42]{Aulakh:2005}:
\be
D_{ij} =
\begin{array}{r l}
0 & \mbox{$i \neq j$} \\
-1 & \mbox{$i = j = 2, 3, 6, 7$} \\
1 & \mbox{$i = j = 1, 4, 5, 8, 9, 0$}.
\end{array}
\ee
To see the detail, we note that the D-parity flipped vevs are  as follows:
\begin{equation}
\label{eq:shiftedvev}
\begin{array}{c}
\Phi_{1234} = \Phi_{1256} = \Phi_{3456} = a, \\
\Phi_{1278} = -\Phi_{1290} = \Phi_{3478} =
-\Phi_{3490} = \Phi_{5678} = -\Phi_{5690} = \omega, \\
\Phi_{7890} = -p,\\
\\
\Sigma_{a+1,b+3,c+5,d+7,e+9} = i^{-a-b-c+d-e} \frac{\sigma}{2^{5/2}},
~~~~ a, b, c, d, e \in \{0,1\}, \\
\
\bSigma_{a+1,b+3,c+5,d+7,e+9} = i^{a+b+c-d+e} \frac{\bsigma}{2^{5/2}},
~~~~ a, b, c, d, e \in \{0,1\}. 
\end{array}
\end{equation}

On the other hand the LR-parity flipped vevs are :
\begin{equation}
\label{eq:LRvev}
\begin{array}{c}
\Phi_{1234} = \Phi_{1256} = \Phi_{3456} = a, \\
\Phi_{1278} = \Phi_{3478} = \Phi_{5678} = 
-\Phi_{129\,10} = -\Phi_{349\,10} = -\Phi_{569\,10} = \omega, \\
\Phi_{789 10} = -p,\\
\\
\Sigma_{a+1,b+3,c+5,d+7,e+9} = i^{a+b+c-d+e} \frac{\sigma}{2^{5/2}},
~~~~ a, b, c, d, e \in \{0,1\}, \\
\\
\bSigma_{a+1,b+3,c+5,d+7,e+9} = i^{-a-b-c+d-e} \frac{\bsigma}{2^{5/2}},
~~~~ a, b, c, d, e \in \{0,1\}. 
\end{array}
\end{equation}
Observe that 
\[
D (\langle 210 \rangle) =
(L \leftrightarrow R) (\langle 210 \rangle),
D (\langle 126 \rangle) =
\frac{\sigma}{\bsigma} 
(L \leftrightarrow R) (\langle \overline{126} \rangle),
D (\langle \overline{126} \rangle) =
\frac{\bsigma}{\sigma} 
(L \leftrightarrow R) (\langle 126 \rangle).
\]
Thus D-parity mimics left right flipping in $\SO(10)$ MSGUT. 

While this discrete symmetry
is securely embedded in a compact group, the engineering of the superpotential  required to
obtain the MSSM encodes an accidental symmetry into the $F$ flatness conditions according to which
for every choice of vev's resulting in MSSM, there exists a set of $D$-flipped vev's which satisfy the
same $F$ flatness conditions. Further, as argued in \cite{Garg:2018}, there exist no flat directions connecting these
mutually flipped set of vev's. On the other hand a one parameter curve $U(1)_D$ generated by $D$
necessarily crosses a barrier in connecting the two sets of vev's. Thus the vacuum manifold possesses accidental
 discrete symmetry, which can give rise to pseudo-topological defects due to causal structure of the evolving
 early Universe. Due to rapid cooling of the universe, such walls can remain metastable until destroyed by
 fluctuation and tunneling processes. The situation becomes analogous to transient DW  arising from breaking of 
 LR symmetry in certain SUSYLR models \cite{Mishra:2010}, although the walls are only metastable. 
In the LR case it was possible to assume Planck scale suppressed terms that break the discrete LR symmetry
since gravity does not respect global symmetries. For a compact group $Spin(10)$ there is no consistent way
to generate an energy imbalance between sectors related by a discrete symmetry operation which belongs to the 
group. This is why we shall need to investigate other methods for domain
wall removal under the action of D-parity in $\SO(10)$ MSGUT.

We start this investigation by computing the F-terms evaluated at the original vevs as well 
as at the flipped vevs. D-flatness at the original and flipped vevs is ensured by utilising the condition
$|\sigma| = |\bsigma|$ as in \cite{Bajc:2004}. The evaluations at the original vevs can also be
found in \cite[Equations~6-9]{Bajc:2004} and \cite[Equations~20-23]{AulakhGirdhar:2005}. 
Now we note that exactly the same  values of $a$, $\omega$, $p$ ensure F-flatness
at the original vevs as well as at the $D$-flipped vevs. The detail is shown in Table \ref{tab:dflip}.
\begin{table}[tbh]
\begin{centering}
\begin{tabular}{| l  c | l | l |}
 \hline
 \hline 
& & & \\
F-term  & \  & Original vev & $D$-Flipped vev \\
& & & \\
\hline 
& & & \\
$
F_{\Phi_{1234}} =
F_{\Phi_{1256}} =
F_{\Phi_{3456}} 
$ 
& = 
&
$
2 m a + 2 \lambda(a^2 + 2 \omega^2) + \eta \sigma \bsigma
$
&
$
2 m a + 2 \lambda(a^2 + 2 \omega^2) + \eta \sigma \bsigma 
$, \\
& & & \\
$
F_{\Phi_{1278}} =
F_{\Phi_{3479}} =
F_{\Phi_{5678}} 
$ 
& = 
&
$
2 m \omega + 2 \lambda (2 a + p) \omega - \eta \sigma \bsigma 
$
&
$
2 m \omega + 2 \lambda (2 a + p) \omega - \eta \sigma \bsigma 
$, \\
& & & \\
$
F_{\Phi_{1290}} =
F_{\Phi_{3490}} =
F_{\Phi_{5690}} 
$ 
& = 
&
$
2 m \omega + 2 \lambda (2 a + p) \omega - \eta \sigma \bsigma 
$
&
$
-2 m \omega - 2 \lambda (2 a + p) \omega + \eta \sigma \bsigma 
$, \\
& & & \\
$
F_{\Phi_{7890}} 
$
& = 
&
$
2 m p + 6 \lambda \omega^2 + \eta \sigma \bsigma 
$
&
$
-2 m p - 6 \lambda \omega^2 - \eta \sigma \bsigma 
$ \\
& & & \\
Other
$
F_{\Phi_{ijkl}}
$
& = 
&
$
0
$
&
$
0
$ \\
& & & \\
$
|F_{\Sigma_{a+1,b+3,c+5,d+7,e+9}}|
$
& = 
&
$
\frac{|\bsigma|}{2^{3/2}}|M + \eta(3a + p - 6\omega)|
$
&
$
\frac{|\bsigma|}{2^{3/2}}|M + \eta(3a + p - 6\omega)|
$ \\
& & & \\
$
|F_{\bSigma_{a+1,b+3,c+5,d+7,e+9}}|
$
& = 
&
$
\frac{|\sigma|}{2^{3/2}}|M + \eta(3a + p - 6\omega)|
$
&
$
\frac{|\sigma|}{2^{3/2}}|M + \eta(3a + p - 6\omega)|
$ \\
& & & \\
Other
$
F_{\Sigma_{ijklm}},
F_{\bSigma_{ijklm}}
$
& = 
&
$
0
$
&
$
0
$ \\
& & & \\
\hline
\end{tabular}
\end{centering}
\caption{Property of various $F$-terms under $D$-parity flip
}
\label{tab:dflip}
\end{table}
From  table \ref{tab:dflip} we see that the $F$-terms evaluated at the original
and $D$-parity flipped vevs are either the same, if the field value 
remains same
after shifting, or negated if the field value gets negated after flipping.
This is not a coincidence, but rather a consequence
of the $SO(10)$-invariance of the superpotential where
each index occurs exactly twice in a term. 
This is true even if we add higher degree
$\SO(10)$ invariant non-renormalisable terms to the superpotential.
So if $|\sigma| = |\bsigma|$, the
scalar potential is the same whether evaluated at the original or the
D-parity
flipped vevs. Thus, a $\SO(10)$-invariant superpotential will never 
give a pressure difference. It becomes necessary to consider other alternatives.

\section{Non-renormalisable superpotential term} 
\label{sec:nonrenormalisable}
Since perfect $\SO(10)$ symmetry in the superpotential at and above 
GUT scale cannot destablise
the non-topological domain wall, we need a `mild' breaking of 
$\SO(10)$ symmetry.
We are thus led to consider adding Planck suppressed degree four terms 
to the superpotential
that depart from $\SO(10)$ invariance. Clearly, this must be
done cautiously in order to affect only the UV end of the theory 
without  sacrificing the desirable features of the
low energy effective renormalisable theory.
We do not speculate on the nature of the physics 
responsible for these higher order terms, but instead adopt a 
`minimal' approach. Besides ensuring removal of the non-topological
domain wall in the early universe, it will be
nice if this mild SO(10) breaking provides some explanation also for
the eventual disappearance of D-parity flipped particle physics in the low energy Universe. 
As will become clear below, our proposal for the  non-renormalisable terms 
give  the D-parity flipped vev slightly higher energy making them disfavoured.
We show in this section how a `minimal' $\SO(10)$ violating but SM
preserving scenario can 
be systematically constructed as a generalisation of the Gell-Mann-Okubo 
``octet dominance hypothesis" from the eightfold way of flavour $SU(3)$. 
Indeed we take a more drastic step of
adapting this to a true gauge symmetry. 
Thus, our full superpotential
now mildly breaks SO(10) but still does not jeopardise SM.

\subsection{Gell-Mann Okubo Formalism}
\label{sec:GMO}
We briefly recall the formalism of Gell-Mann \cite{GellMann:1961} and Okubo \cite{Okubo:1962} 
who independently gave an explanation for the mass splittings within several of the multiplets of flavour $SU(3)$, viz., 
the pseudoscalar mesons octet, baryon octet and baryon decuplet.  (For the later more systematic 
developments see \cite{Gasser:1980sb,Jenkins:1991ts,Banerjee:1994bk}).
Suppose one starts with exact flavour symmetry amongst the
$u$, $d$ and $s$ quarks. In other words, we consider the 
global symmetry group $\SU(3)$. The pseudoscalar mesons are made up of these
three quarks and can be arranged to form an octet $\Pi$ in the adjoint 
representation of $\SU(3)$. The Hamiltonian describing them at rest
can be written as
$
H_0 = \frac{\mu^2}{2} \Tr [\Pi^\dag \Pi],
$
where $\mu^2$ denotes their squared mass. This Hamiltonian is 
$\SU(3)$ symmetric and predicts that all the pseudoscalar mesons
will have exactly the same mass $\mu$. However, this is not the case.
This deviation from universal masses can be explained by postulating a perturbed Hamiltonian
$
H = H_0 + H',
$
where the perturbation $H'$ is given by
$
H' = \frac{\alpha}{2} \Tr [\Phi^\dag \Phi M].
$
The $3 \times 3$ `coefficient matrix' $M$ is chosen so as to break
$\SU(3)$ symmetry but preserve the lower energy isospin $\SU(2)$
symmetry. This is done by taking $M$ from the Lie algebra $\su(3)$ 
and requiring it to commute with $\su(2)$ embedded
into $\su(3)$. This requirement fixes $M$ uniquely to be the
Gell-Mann $\lambda_8$. 

\subsection{Choice of $SO(10)$ breaking matrix}
\label{sec:breaking}
Along similar lines we now consider the following extended superpotential:
\be
W = 
W_{\mathrm{ren}} +
\frac{b}{M_{\mathrm{Pl}}} W_{\mathrm{nr}},
\ee
where 
\be
W_{\mathrm{nr}} =
\frac{1}{(4!)^4} (\Phi^T M' \Phi)^2.
\ee
Above $M'$ is a linear operator acting on the column vector $\Phi$. 
We shall take $M'$ to be the representation
of a carefully chosen group element $M$ of $\SO(10)$. With a matrix $M$ in the notation
of Sec. \ref{sec:degeneratevacua},   
$M' = M^{\otimes 4}$. We require that $M$
commute with all group elements $N \in \SO(10)$ that arise as images of
the embedding of the SM group $\SU(3)_c \times \SU(2)_L \times \U(1)_Y$ into
$\SO(10)$. 
It is then clear that $W_{\mathrm{nr}}$ 
is SM-invariant
but not necessarily $\SO(10)$ invariant. 

We can now utilise the Pati-Salam embedding of SM group into $\SO(10)$ as given 
in \cite[Section~2.2]{Aulakh:2005}. It is easy to see that $\SO(10)$  matrices of the form 
\be
M = 
\left(
\begin{array}{c | c}
\one_{6 \times 6} & 0_{6 \times 4} \\
\hline \\
0_{4 \times 6} & K_{4 \times 4}
\end{array}
\right)
\ee
commute with all $\SO(10)$ matrices $N$ that arise from the embedding
of the SM group into $\SO(10)$. 
Above $K$ is an $\SO(4)$ matrix of the
form $K = \exp(2 \theta J)$, where 
\be
J = 
j_1 J_1^+ + j_2 J_2^+ + j_3 J_3^+ \in \so(4),
\ee
$\theta$ is a real number between $0$ and $2\pi$,
$j_1$, $j_2$, $j_3$ are real numbers satisfying
$j_1^2 + j_2^2 + j_3^2 = 1$, and
\be
J_1^+ = 
\frac{1}{2}
\left(
\begin{array}{c c c c}
0 &  0 &  0 &  1 \\
0 &  0 &  1 &  0 \\
0 & -1 &  0 &  0 \\
-1 &  0 &  0 &  0
\end{array}
\right),
J_2^+ = 
-\frac{1}{2}
\left(
\begin{array}{c c c c}
0 &  0 &  1 &  0 \\
0 &  0 &  0 & -1 \\
-1 &  0 &  0 &  0 \\
0 &  1 &  0 &  0
\end{array}
\right),
J_3^+ = 
\frac{1}{2}
\left(
\begin{array}{c c c c}
0 &  1 &  0 &  0 \\
-1 &  0 &  0 &  0 \\
0 &  0 &  0 &  1 \\
0 &  0 & -1 &  0
\end{array}
\right)
\ee
are the three self-dual generators of $\so(4)$. Recall that the
self-dual elements of $\so(4)$ form the embedding of $\su(2)_R$ into
$\so(4)$. It can now be shown that $K$ is the image of the 
matrix 
$\exp(i \theta (j_1 \tau_1 + j_2 \tau_2 + j_3 \tau_3)) \in \SU(2)_R$
when embedded into $\SO(4)$.  

The reason behind $M$ commuting with SM is as follows.
Recall that $\SU(3)_c$ of
SM gets embedded into the $\SO(6)$ part on the top left. 
The generator of $u(1)_{B-L}$ then maps to the $\so(10)$ matrix
\be
B-L=\left(
\begin{array}{c | c}
\begin{array}{r r r r r r}
0  & 1 & 0 & 0 &  0& 0\\
-1 & 0 & 0& 0& 0& 0\\
 0 & 0&  0 & 1 & 0& 0\\
0  & 0& -1 & 0 & 0& 0\\
0  & 0& 0& 0&  0 &  1  \\
 0 & 0& 0& 0& -1 & 0  
\end{array}
& O_{6 \times 4} \\
\hline
O_{4 \times 6} 
& O_{4 \times 4} 
\end{array}
\right).
\ee
The $\su(2)_L$ of SM maps into the 
anti-self-dual generators and $\su(2)_R$ maps
into the self-dual generators 
of $\so(4)$ in the bottom right.
Since the self-dual generators of $\so(4)$ commute with the 
anti-self-dual generators, any linear
combination of the three self-dual generators of $\so(4)$ will
commute with $\su(2)_L$. We fix one such linear combination
$
J = 
j_1 J_1^+ + j_2 J_2^+ + j_3 J_3^+ \in \so(4),
$
and map $\mathrm{u}(1)_R$ to it. 
This $J$ commutes with the embedding of
$
\su(3)_c \times \mathrm{u}(1)_{B-L} 
\times \su(2)_L \times \mathrm{u}(1)_R
$
into $\so(10)$ and thus with 
the embedding of SM into $\so(10)$.

We shall take $\theta = \pi/2$, $j_1 = j_3 = \frac{1}{\sqrt{2}}$, 
$j_2 = 0$ to get
\be
K = 
\frac{1}{\sqrt{2}}
\left(
\begin{array}{c c c c}
0 &  1 &  0 &  1 \\
-1 &  0 &  1 &  0 \\
0 & -1 &  0 &  1 \\
-1 &  0 & -1 &  0
\end{array}
\right).
\ee
This gives us
\be
M =
\frac{1}{\sqrt{2}}
\left(
\begin{array}{c | c}
\sqrt{2} \, \one_{6 \times 6} & 0_{6 \times 4} \\
\hline \\
0_{4 \times 6} & 
\begin{array}{c c c c}
0 &  1 &  0 &  1 \\
-1 &  0 &  1 &  0 \\
0 & -1 &  0 &  1 \\
-1 &  0 & -1 &  0
\end{array}
\end{array}
\right).
\ee

Observe that the element $M \in \SO(10)$ does not commute with D-parity.
In other words, the D-parity generator viz. 
\be
J_{23} + J_{67} =
\left(
\begin{array}{r r r r r r r r r r}
0    &  0 &  0 & 0 &  0 & 0  & 0 & 0 & 0 & 0 \\
  0   &  0 &  1 & 0  & 0   &  0  & 0  &  0 &  0 & 0\\
 0  & -1 &  0 & 0  &  0  &  0  &  0 & 0  & 0  & 0\\
0    &  0 &  0 & 0 &  0 & 0  & 0 & 0 & 0 & 0 \\
0    &  0 &  0 & 0 &  0 & 0  & 0 & 0 & 0 & 0 \\
 0    & 0   &  0  &  0 &  0  &  0 & 1 & 0  & 0  & 0\\
 0    & 0   & 0   & 0  & 0   & -1 & 0 & 0  & 0  & 0\\
0    &  0 &  0 & 0 &  0 & 0  & 0 & 0 & 0 & 0 \\
0    &  0 &  0 & 0 &  0 & 0  & 0 & 0 & 0 & 0 \\
0    &  0 &  0 & 0 &  0 & 0  & 0 & 0 & 0 & 0 
\end{array}
\right)
\ee
is a broken symmetry generator in the non-renormalisable superpotential.
This allows us the possibility of breaking the degeneracy between the
two sets of vevs by means of the non-renormalisable term 
$W_{\mathrm{nr}}$ in the superpotential.  If there exist basic reasons for the existence of such a term
then there could be more terms of the same type of higher mass dimensions. However this leading term
should suffice for our purpose.

\subsection{The effective potential and domain wall removal}
\label{sec:effectivepotential}
We next proceed to compute the effective potential in the two quasi symmetric
vacua now split by the new non-renormalisable term. 
Observe that
\be
\frac{\partial W_{\mathrm{nr}}}{\partial \Phi_{ijkl}} =
\frac{2 (\Phi^T M^{\otimes 4} \Phi)}
     {(4!)^4}
\frac{\partial (\Phi^T M^{\otimes 4} \Phi)}{\partial \Phi_{ijkl}},
\ee
To keep the computation simple 
we note that the vev's themselves will shift by $O(1/M_{\mathrm{Pl}})$, and thus
the contribution to the effective potential from the change in the vev's is  $O(1/M^4_{\mathrm{Pl}})$ 
and can be ignored compared to the difference of $O(1/M^2_{\mathrm{Pl}})$ arising from the leading Planck suppressed
terms. Thus we can use the vev's already determined by Bajc et al. \cite{Bajc:2004}.
In appendix \ref{sec:appA} we have shown the evaluation of the resulting effective potential in this approximation.
Since the scalar potentials evaluated at the $D$-flipped vevs are not the
same, we get a pressure difference across the domain wall as follows
\be
\begin{array}{rcl}
\lefteqn{
\left. V \right|_{\mbox{flipped vevs}} -
\left. V \right|_{\mbox{original vevs}} 
} \\
&  & \\
& = &
\frac{b^2}{M_{\mathrm{Pl}}^2}
(4^2 (3 a^2 + 6 \omega^2 + p^2))
(
(3 a^2 + 6 \omega^2 + p^2)^2
-
(3 a^2 + p^2)^2
).
\end{array}
\ee

From the above expression for the pressure difference, it is clear
that $\omega$ is the controlling parameter for the difference in the
scalar potential between the original and flipped vevs. For
most `interesting' regions of the parameter space,
$\max\{a, p\} \gg \omega$ 
\cite{Bajc:2004, Aulakh:2004, AulakhGirdhar:2005}. For example, 
consider some `representative' values 
\[
a = -0.67 \frac{m}{\lambda}, ~~~
\omega = -0.21 \frac{m}{\lambda}, ~~~
p = -0.27 \frac{m}{\lambda}, ~~~
\sigma = \bsigma = 0.51 \frac{m}{\sqrt{\eta \lambda}}, ~~~
\lambda = 0.12, ~~~
\eta = 0.21,
\]
taken from Equation~4 of Aulakh et al. \cite{Aulakh:2004}  and 
Equation~57 of Aulakh and
Girdhar \cite{AulakhGirdhar:2005}.
This corresponds to
$x = 0.21$ and $\frac{\lambda M}{\eta m} = 1.03$ 
in Equation~4 of  \cite{Aulakh:2004}.
The mass $m$ is set to the mass $M_X$ which is the lightest 
superheavy
vector particle mediating proton decay, as mentioned in page~291 of
\cite{AulakhGirdhar:2005}. For the above parameters,
we take $m = 10^{14}~\mathrm{GeV}/c^2$ using
Fig.~4 and Equation~55 of Aulakh and Girdhar \cite{AulakhGirdhar:2005}.
This gives us in units of $\mathrm{GeV}/c^2$
\[
a \sim -10^{15}, ~~~
\omega \sim -10^{14}, ~~~
p \sim -10^{14}, ~~~
\sigma = \bsigma \sim 10^{15},
\]
leading to $a \gg p \geq \omega$.
With these vevs, $\SO(10)$ MSGUT breaks down directly to MSSM.
The assumption $\max\{a, p\} \gg \omega$ occurs for many other 
regions of parameter space
breaking $\SO(10)$ MSGUT directly to MSSM, as well as 
for the case of one or
two intermediate scales of symmetry breaking with both Pati-Salam
and left-right symmetric MSSM as the intermediate gauge groups
\cite[Section~V]{Bajc:2004}.

In the following, 
we shall take $\max\{a, p\} \sim M_X$ and $\omega \sim M_R$, 
where $M_R$ is the (lower) energy scale at which  a scalar field 
acquires a non-zero vev also setting the scale of domain wall tension. 
With this, the pressure difference becomes
\begin{equation}
\label{eq:pressurediff}
|\left. V \right|_{\mbox{flipped vevs}} -
 \left. V \right|_{\mbox{orig. vevs}}| \sim
\frac{b^2}{M_{\mathrm{Pl}}^2} M_X^4 M_R^2.
\end{equation}

\section{Constraint from cosmology}
\label{sec:DWremoval}
We now derive the constraint the parameter $b$ must satisfy if
the non-renormalisable term were to be able to
remove any domain walls that may have arisen in the course of $SO(10)$ breaking.
Following Mishra and Yajnik \cite{Mishra:2010}, we consider the following three eras for domain wall 
appearance and removal. 

\subsubsection{Evolution in radiation dominated universe}

For most grand unified theories, barring the inflationary period, the universe is in a thermal state,
and radiation dominated. The formation and evolution of domain walls in such a medium
is a difficult problem, requiring numerical simulations. However, strong analytical arguments
have been given and the essentials of this scenario was originally proposed by Kibble
\cite{Kibble:1980mv} and Vilenkin \cite{Vilenkin:1984ib}. We refer the reader to the original papers,
or to a recapitulation in \cite{Mishra:2010}.  Suppose the Domain walls 
arise at some temperature  $T_c$, the critical temperature of a phase transition at which a scalar field 
$\phi$ acquires a non-zero vacuum expectation value. We shall be primarily interested in this scale 
being $M_R$, being the scale after which the Universe could accidentally have fallen into a $D$-flipped
vacuum with $SU(2)_L$ of electroweak replaced by $\rightarrow$ $SU(2)_R$. The energy 
density trapped per unit area of the wall is $\sigma \sim M_R^3$. 
One can then argue that the required pressure across the walls to destabilise them must be 
\begin{eqnarray}
 \delta \rho &\ge& G \sigma^2 \\
 &\approx& \frac{M_R^6}{M_{Pl}^2}
\label{eq:eps-vsix}
\end{eqnarray}
where the second equation is obtained by substituting the relevant scales.
combining this with Eqn.~(\ref{eq:pressurediff}) gives
\be
b^2 \geq \frac{M_R^4}{M_X^4} \qquad \mathrm{Radiation\ dominated\ case}. 
\ee

\subsubsection{Evolution in matter dominated universe}
The next two possibilities arise naturally in the context of removal moduli in 
superstring cosmology \cite{Coughlan:1983ci,Banks:1993en,deCarlos:1993jw} (see \cite{Kane:2015jia} for a recent review).
This issue has received fresh attention in \cite{Dutta:2014tya}\cite{Cicoli:2016olq}\cite{Bhattacharya:2017pws}\cite{Bhattacharya:2017ysa}
\cite{Goswami:xyz}, 
in view of the improved inputs from Cosmic Microwave Background data. 
When a modulus field begins to oscillate coherently
it can
be seen that near the minimum of the potential the oscillations are essentially harmonic, 
$\bar{KE}=\bar{PE}$, ie, making $p =T^i_i= KE-PE\approx 0$ where there is no summation
on the generic space component index $i$ of the energy-momentum tensor. 
This is equivalent to a matter dominated era by the time the domain walls get destabilised. 

The simpler possibility within this scenario is that the domain walls start decaying as soon as
they dominate the energy density of the Universe. Thus no competition ensues between them and the 
oscillating moduli.
This possibility was considered in \cite{Kawasaki:2004rx} by Kawasaki and Takahashi.
The analysis begins by assuming that the initially formed wall complex 
in a phase transition rapidly relaxes to a few walls per horizon 
volume at an epoch characterized by Hubble parameter value $H_i$. 
Due to the assumption of almost simultaneous onset of degradation of domain walls,
if the temperature at this particular epoch is $T_D$, then $H_{eq}^2 \sim G T_D^4$ where
\begin{equation}
H_{eq} \sim \sigma^{\frac{3}{4}} H_i^{\frac{1}{4}} M_{Pl}^{-\frac{3}{2}} ~,
\label{eq:Heq}
\end{equation}
from 
which we find that
\begin{equation}
 T_D^4 \sim \sigma^{\frac{3}{2}}H_i^{\frac{1}{2}}M_{Pl}^{-1}
\label{eq:TD4}
\end{equation}
Putting together Eq.s ({eq:Heq}) and (\ref{eq:TD4}) we get,
\begin{equation}
 T_D^4 \sim \frac{\sigma^{11/6}}{M_{Pl}^{3/2}} 
\label{eq:TD4I}
\end{equation}
Now requiring $\delta\rho > T_D^4$ and substituting for $\sigma$ we get,
\begin{equation}
\delta\rho> M_R^4 \left(\frac{M_R}{M_{Pl}}\right)^{3/2}
\label{eq:eps-v11half}
\end{equation}
Consequently, for $b$ this implies,
\be
b^2 \geq
\frac{M_R^{7/2} M_{\mathrm{Pl}}^{1/2}}{M_X^4}. \qquad \mathrm{moduli\ dominated\ case}
\ee

\subsubsection{Evolution including weak inflation}
The third possibility we consider is that both moduli fields as well the walls are present, 
and unlike the simplified case above, the latter do not disintegrate by
the time they come to dominate the energy density of the Universe.
 A late period of exponential growth of the scale factor can occur due to some moduli frozen
 at zero value due to thermal corrections, later relaxing to their true minimum, when
 the universe also mildly reheats.
 This possibility, weak or thermal inflation,  was considered \cite{Lyth:1995hj},\cite{Lyth:1995ka}, with some recent
developments reported in \cite{Hong:2015oqa,Hayakawa:2015fga,Cho:2017zkj}.
During the epoch when the weak inflation has ended and the moduli have begun to oscillate,
they obey a matter equation of state.
But the equation of state of for non-scaling domain walls is $p=-(2/3)\rho$, and the Friedmann 
scale factor evolves as $a(t) \propto t^2$.
Then the walls dominate over the coherent moduli fields.
Clearly this domination must last only for a limited epoch to result in acceptable cosmology. 

Wall destabilisation mechanism in this case has not been widely pursued, and hence 
we recapitulate in detail the  original reasoning of \cite{Mishra:2010}.
The given situation is most likely in the case when the
$\delta \rho$ is typically small, not large enough
to destabilize the walls sufficiently quickly. The two kinds of dynamics interfering with the
decay mechanism are rapid oscillations with large tension in the walls, of rapid bulk 
sweeping and friction with the medium. But eventually a small
pressure difference will also win over the tension force 
because  the walls straighten out, 
or the frictional force as the translational speed reduces
drastically. Since we have no microscopic model for deciding which
of these is finally responsible, we introduce a temperature scale
$T_D$ at which the walls begin to experience instability.
Note that unlike in the previous example, we will not be able to
estimate $T_D$ in terms of other mass scales and will accept it as 
undetermined and consider a few reasonable values for it for our final
estimate.

As has been studied above, at $H_{eq}$ the energy density of the domain
wall network dominates energy density of the Universe.
The scale factor at this  epoch is characterized by
$a_{eq}$. Denoting the energy density of the domain walls
at the time of equality as $\rho_{DW}(t_{eq})$,
the evolution of energy density can be written as,
\begin{equation}
 \rho_{DW}(t_d)\sim \rho_{DW}(t_{eq})\left( \frac{a_{eq}}{a_d}\right)
\label{eq:rhodw}
\end{equation}
where $a_d$ is scale factor at the epoch of decay of domain
wall corresponding to time $t_d$. If the domain walls decay at an
epoch characterized by temperature $T_D$, then $\rho_{DW}(t_d)\sim T_D^4$.
So from Eq.(\ref{eq:rhodw}),
\begin{equation}
 T_D^4=
 \rho_{DW}(t_{eq})\left( \frac{a_{eq}}{a_d}\right)
\label{eq:rhoeq}
\end{equation}

In the matter dominated era the energy density of the moduli
fields scale as,
\begin{equation}
 \rho^d_{mod} \sim \rho^{eq}_{mod}\left( \frac{a_{eq}}{a_d}\right)^3
\label{eq:rhoMod}
\end{equation}
Substituting the value of $a_{eq}/a_d$ from Eq.(\ref{eq:rhoeq}) in
the above equation,
\begin{equation}
 \rho^d_{mod} \sim \frac{T_{D}^{12}}{\rho^2_{DW}(t_{eq})}
\label{eq:rhoModTD}
\end{equation}
Since the energy density of the domain walls dominates
the universe after the time of equality,
$\rho_{DW}(t_d) > \rho^d_{mod}$. So the pressure
difference across the domain walls when they start
decaying is given by,
\begin{equation}
 \delta \rho \gtrsim \frac{T_D^{12} G^2}{H_{eq}^4}
\label{eq:deltaRho}
\end{equation}
where we have used the relation $H_{eq}^2 \sim G \rho_{DW}(t_{eq})$.
Replacing the value of $H_{eq}$ from Eq.(\ref{eq:Heq}),  and 
$H_i^2 \sim G \rho^{in}_{DW}\sim M_R^4/M_{Pl}^2$, 
\begin{equation}
 \delta \rho \gtrsim M_R^4 \left(\frac{T_D^{12} M_{Pl}^3}
{M_R^{15}}\right)
\label{eq:deltaRhoTD}
\end{equation}
For the case of weak inflation, 
Equations~(\ref{eq:pressurediff}) and (\ref{eq:deltaRhoTD}) lead to the following constraint
on the parameter $b$
\be
|b|^2 \geq
\frac{T_D^{12} M_{\mathrm{Pl}}^{5}}{M_R^{13} M_X^4} \qquad \mathrm{Weak\ inflation\ with\ moduli}
\ee

\subsection{Constraints on $b$}
This completes the discussion of the three scenarios of wall degradation. 
We now tabulate the constraints arising from the above
inequalities, assuming $M_X = 10^{16}$ viz. of the order of
the GUT scale, and
$M_{Pl} = 10^{19}$. We consider three candidate values for 
$M_R$: a `low' value of $10^{7}$ corresponding to non-thermal or
resonant leptogenesis, an `intermediate' value of
$10^9$ consistent with thermal leptogenesis without gravitino
overabundance, and a `large' value of $10^{13}$. 
For the radiation dominated and matter dominated eras, we obtain
constraints on $|b|$ in order to ensure successful domain wall removal. 
For the case
of weak inflation, we assume $|b| \leq O(1)$ and instead calculate
constraints on the ratio $T_D / M_R$ in order to ensure successful
domain wall removal. This is because the energy scale $T_D$ where the
walls first experience instability is unknown to us.

\begin{table}[tbh]
\begin{minipage}[t]{0.45\textwidth}
\begin{center}
\begin{tabular}{c | c | c}
Era & $M_R$     & $b \geq {}$  \\
\hline 
    & $10^7$    & $10^{-18}$   \\
RD  & $10^9$    & $10^{-14}$   \\
    & $10^{13}$ & $10^{-6}$   \\
\hline                         
    & $10^7$    & $10^{-15}$   \\
MD  & $10^9$    & $10^{-11.5}$ \\
    & $10^{13}$ & $10^{-4.5}$ 
\end{tabular}
\end{center}
\end{minipage}
~
\begin{minipage}[t]{0.45\textwidth}
\begin{center}
\begin{tabular}{c | c | c}
Era & $M_R$     & $b \leq 1, \frac{T_D}{M_R} \leq {}$ \\
\hline 
    & $10^7$    & $10^{-3.17}$   \\
WI  & $10^9$    & $10^{-3.3}$   \\
    & $10^{13}$ & $10^{-3.67}$  
\end{tabular}
\end{center}
\end{minipage}
\caption{
Constraints arising from different values of $M_R$ in different
cosmological eras
}
\label{tab:constraints}
\end{table}

\smallskip

From Table~\ref{tab:constraints}, we see that
our model is easily capable of ensuring
domain wall removal in the radiation dominated and matter dominated
eras without conflicting
with existing experimental data. This is because the constraints on
$|b|$ are very easy to meet; the minuscule symmetry breaking term
is enough to cause instability in domain walls formed during the two
eras. For the case of weak inflation, assuming $|b| \leq O(1)$, we
get constraints on the ratio $T_D / M_R$. The ratio is required to be
less than $10^{-3}$. For $T_D>10$MeV as required by BBN, this comfortably 
accommodates the $\gtrsim 10^{10}$GeV scale expected of thermal leptogenesis.
However an $M_R$ scale of $10 - 100$TeV scale that may be pursued at future colliders
is marginally acceptable.

\section{Conclusion}
\label{sec:conclusion}
In this paper, we take the $\SO(10)$ based MSGUT studied extensively
in earlier works and examine it from a cosmological angle. We have identified the occurrence of
topological pseudo-defects which arise from applying the D-parity operator to the SUSY preserving 
vevs breaking $\SO(10)$ MSGUT.  In addition to the intended left handed MSSM  this operation 
gives a new  set of SUSY preserving vevs exchanging $SU(2)_L$ with $SU(2)_R$. These two sets 
of vevs are disconnected in the sense that there is no SUSY preserving path connecting them in
parameter space. We then remark that this leads to a danger of long lived metastable domain
wall formation in the early universe especially close to the Planck scale. A similar conclusion has 
recently been made in \cite{Garg:2018}. 

We then investigate conditions under which domain walls, if formed,
can go away early enough so as not to conflict with standard cosmology.
We first observe that domain walls
cannot go away even with dimension four operators in the superpotential
as long as the theory is $\SO(10)$ symmetric.
We then postulate a non-renormalisable
term that breaks of $\SO(10)$ symmetry while still preserving SM 
symmetry. 
While we are unable to argue why Planck scale effects should distinguish 
the two mirror image subgroups, we have proposed the least disruptive 
way of avoiding undesirable cosmological consequences, and keeping in 
mind the open ended nature of UV completion of known physics, we could 
say the theory is unified up to Planck scale effects, or only quasi-unified. 
Armed with this, we investigate domain wall removal in three
scenarios of the early universe viz. radiation dominated era,
matter dominated era which is generic to string theory inspired models,
and a weak inflation era following matter domination. 

Our main conclusions are as follows:
\begin{enumerate}

\item
If domain walls are formed in the radiation dominated era, they can
easily disappear in that era itself in our model. This is true for
a wide range of energy scales of domain wall formation. The same is true
if domain walls are formed within the epoch dominated by moduli which 
make the universe approximately matter dominated. They can then disappear
promptly with aid from the softly $SO(10)$  breaking operators. 
The parameter characterising the non-renormalisable term is not constrained 
for a wide range of values of the domain wall energy scale in both of these scenarios.

\item 
If domain walls are formed within the moduli dominated era and continue
on during the coherent oscillation phase weak inflation, then domain walls can eventually
completely disappear only under a certain condition. The condition 
required by
our model is that the temperature at which the walls become unstable 
be at least three to four orders of magnitude smaller than the energy 
density at which the walls first appear. This still allows consistency with
thermal leptogenesis subject to tackling gravitino overabundance. However a $10 - 100$ TeV
scale right handed sector becomes only marginally acceptable.

\end{enumerate}

Thus this work invites further careful analysis of other 
cosmological implications of $\SO(10)$  MSGUT.

\appendix
\section{}
\label{sec:appA}
We begin by evaluating the new term, expressing it in terms of  components,
\be
\begin{array}{rcl}
\lefteqn{
\frac{1}{(4!)^2}
\Phi^T M^{\otimes 4} \Phi 
} \\
& = &
\displaystyle \sum_{i<j<k<l \leq 6} \Phi_{ijkl}^2 \\
&   &
{} +
\displaystyle \sum_{i<j \leq 6} 
\left(
\frac{\Phi_{ij78}^2}{2} +
\Phi_{ij78} \Phi_{ij70} +
\Phi_{ij78} \Phi_{ij89} -
\Phi_{ij78} \Phi_{ij90} +
2 \Phi_{ij79} \Phi_{ij80}
\right. \\
&   &
~~~~~~~~~~~~~
\left.
{} +
\frac{\Phi_{ij70}^2}{2} -
\Phi_{ij70} \Phi_{ij89} +
\Phi_{ij70} \Phi_{ij90} +
\frac{\Phi_{ij89}^2}{2} +
\Phi_{ij89} \Phi_{ij90} +
\frac{\Phi_{ij90}^2}{2}
\right) + 
\Phi_{7890}^2.
\end{array}
\ee

Evaluating $W_{\mathrm{nr}}$ at the original vevs, we get
\be
\left.W_{\mathrm{nr}} \right|_{\mbox{orig. vevs}} =
\left(
3 a^2 
+ 3 \left(
\frac{\omega^2}{2} - \omega^2 + \frac{\omega^2}{2}
\right) 
+ p^2
\right)^2 =
(3 a^2 + p^2)^2.
\ee
Evaluating $W_{\mathrm{nr}}$ at the flipped vevs, we get
\be
\left.W_{\mathrm{nr}} \right|_{\mbox{flipped vevs}} =
\left(
3 a^2 
+ 3 \left(
\frac{\omega^2}{2} + \omega^2 + \frac{\omega^2}{2}
\right) 
+ p^2
\right)^2 =
(3 a^2 + 6 \omega^2 + p^2)^2.
\ee

We now evaluate the partial derivatives of 
$
\frac{1}{(4!)^2}
\Phi^T M^{\otimes 4} \Phi
$
with respect to the various fields.
\be
\frac{1}{(4!)^2}
\frac{\partial (\Phi^T M^{\otimes 4} \Phi)}{\partial \Phi_{ijkl}} =
\begin{array}{l l}
2 \Phi_{ijkl} 
& 1 \leq i < j < k < l \leq 6, \\
\Phi_{ij78} + \Phi_{ij70} + \Phi_{ij89} - \Phi_{ij90} 
& 1 \leq i < j \leq 6, kl=78, \\
2 \Phi_{ij80}
& 1 \leq i < j \leq 6, kl=79, \\
\Phi_{ij78} + \Phi_{ij70} - \Phi_{ij89} + \Phi_{ij90} 
& 1 \leq i < j \leq 6, kl=70, \\
\Phi_{ij78} - \Phi_{ij70} + \Phi_{ij89} + \Phi_{ij90} 
& 1 \leq i < j \leq 6, kl=89, \\
2 \Phi_{ij79}
& 1 \leq i < j \leq 6, kl=80, \\
-\Phi_{ij78} + \Phi_{ij70} + \Phi_{ij89} + \Phi_{ij90} 
& 1 \leq i < j \leq 6, kl=90, \\
2 \Phi_{7890}
& ijkl=7890, \\
0 
& \mbox{otherwise}.
\end{array} 
\ee
Evaluating the partial derivatives 
at vevs determined in \cite{Bajc:2004}, we get
\be
\left.
\frac{1}{(4!)^2}
\frac{\partial (\Phi^T M^{\otimes 4} \Phi)}{\partial \Phi_{ijkl}} 
\right|_{\mbox{orig. vevs}} =
\begin{array}{l l}
2 a
& ijkl = 1234, 1256, 3456, \\
2 \omega
& ijkl = 1270, 3470, 5670, 1289, 3489, 5689, \\
2 p
& ijkl=7890, \\
0 
& \mbox{otherwise}.
\end{array} 
\ee
Evaluating the partial derivatives 
at the flipped vevs, we get
\be
\left.
\frac{1}{(4!)^2}
\frac{\partial (\Phi^T M^{\otimes 4} \Phi)}{\partial \Phi_{ijkl}} 
\right|_{\mbox{flipped vevs}} =
\begin{array}{l l}
2 a
& ijkl = 1234, 1256, 3456, \\
2 \omega
& ijkl = 1278, 3478, 5678, \\
-2 \omega
& ijkl = 1290, 3490, 5690, \\
-2 p
& ijkl=7890, \\
0 
& \mbox{otherwise}.
\end{array} 
\ee

We can now compute the contribution to the F-terms arising from 
$W_{\mathrm{nr}}$ for both types of vevs.
\be
\left.
(F_{\mathrm{nr}})_{\Phi_{ijkl}}
\right|_{\mbox{orig. vevs}} =
\begin{array}{l l}
4 (3 a^2 + p^2) a
& ijkl = 1234, 1256, 3456, \\
4 (3 a^2 + p^2) \omega
& ijkl = 1270, 3470, 5670, 1289, 3489, 5689, \\
4 (3 a^2 + p^2) p
& ijkl=7890, \\
0 
& \mbox{otherwise}.
\end{array} 
\ee
\be
\left.
(F_{\mathrm{nr}})_{\Phi_{ijkl}}
\right|_{\mbox{flipped vevs}} =
\begin{array}{l l}
4 (3 a^2 + 6 \omega^2 + p^2) a
& ijkl = 1234, 1256, 3456, \\
4 (3 a^2 + 6 \omega^2 + p^2) \omega
& ijkl = 1278, 3478, 5678, \\
-4 (3 a^2 + 6 \omega^2 + p^2) \omega
& ijkl = 1290, 3490, 5690, \\
-4 (3 a^2 + 6 \omega^2 + p^2) p
& ijkl=7890, \\
0 
& \mbox{otherwise}.
\end{array} 
\ee

Since both the original vevs as well as the flipped vevs are F-flat
at the renormalisable level for suitably chosen values of $a$, $\omega$,
$p$, the above expressions are the complete
values, including renormalisable and non-renormalisable contributions, 
of the respective F-terms. In other words, 
\be
\begin{array}{rcl}
\left. F_{\Phi_{ijkl}} \right|_{\mbox{orig. vevs}}
& = &
\frac{b}{M_{\mathrm{Pl}}} 
\left. (F_{\mathrm{nr}})_{\Phi_{ijkl}} \right|_{\mbox{orig. vevs}}, \\
&  & \\
\left. F_{\Phi_{ijkl}} \right|_{\mbox{flipped vevs}}
& = &
\frac{b}{M_{\mathrm{Pl}}} 
\left. (F_{\mathrm{nr}})_{\Phi_{ijkl}} \right|_{\mbox{flipped vevs}}, \\
&  & \\
\left. F_{\Sigma_{ijklm}} \right|_{\mbox{orig. vevs}}
& = &
\left. F_{\Sigma_{ijklm}} \right|_{\mbox{flipped vevs}}
\;=\;
0.
\end{array}
\ee

The F-term contribution to the scalar potential evaluated
at the original and flipped  vevs becomes
\be
\begin{array}{rcl}
\left. V_F \right|_{\mbox{orig. vevs}}
& = &
\frac{b^2}{M_{\mathrm{Pl}}^2} 
\sum_{k} 
\left. |(F_{\mathrm{nr}})_{f_k}|^2 \right|_{\mbox{orig. vevs}}, \\
&  & \\
\left. V_F \right|_{\mbox{flipped vevs}}
& = &
\frac{b^2}{M_{\mathrm{Pl}}^2} 
\sum_{k} 
\left. |(F_{\mathrm{nr}})_{f_k}|^2 \right|_{\mbox{flipped vevs}}.
\end{array}
\ee
This gives us the following
expressions for F-term contributions to the scalar potential.
\be
\begin{array}{rcl}
\lefteqn{\left. V_F \right|_{\mbox{orig. vevs}}} \\
&   & \\
& = &
\frac{b^2}{M_{\mathrm{Pl}}^2}
(
3 (4 (3 a^2 + p^2) a)^2 +
6 (4 (3 a^2 + p^2) \omega)^2 +
(4 (3 a^2 + p^2) p)^2
), \\
&   & \\
& = &
\frac{b^2}{M_{\mathrm{Pl}}^2}
(4^2 (3 a^2 + p^2)^2 (3 a^2 + 6 \omega^2 + p^2), \\
&  & \\
\lefteqn{\left. V_F \right|_{\mbox{flipped vevs}}} \\
&   & \\
& = &
\frac{b^2}{M_{\mathrm{Pl}}^2}
(
3 (4 (3 a^2 + 6 \omega^2 + p^2) a)^2 +
6 (4 (3 a^2 + 6 \omega^2 + p^2) \omega)^2 +
(4 (3 a^2 + 6 \omega^2 + p^2) p)^2
), \\
&   & \\
& = &
\frac{b^2}{M_{\mathrm{Pl}}^2}
(4^2 (3 a^2 + 6 \omega^2 + p^2)^3).
\end{array}
\ee
Since we take $|\sigma| = |\bsigma|$, both the original and the flipped vevs
are D-flat. Thus, the scalar potentials evaluated at the two types of vevs
arise solely from F-term contributions.


\begin{thebibliography}{39}
\expandafter\ifx\csname natexlab\endcsname\relax\def\natexlab#1{#1}\fi
\expandafter\ifx\csname bibnamefont\endcsname\relax
\def\bibnamefont#1{#1}\fi
\expandafter\ifx\csname bibfnamefont\endcsname\relax
\def\bibfnamefont#1{#1}\fi
\expandafter\ifx\csname citenamefont\endcsname\relax
\def\citenamefont#1{#1}\fi
\expandafter\ifx\csname url\endcsname\relax
\def\url#1{\texttt{#1}}\fi
\expandafter\ifx\csname urlprefix\endcsname\relax\def\urlprefix{URL }\fi
\providecommand{\bibinfo}[2]{#2}
\providecommand{\eprint}[2][]{\url{#2}}

\bibitem{Mohapatra:1980}
\bibinfo{author}{\bibfnamefont{P.}~\bibnamefont{Minkowski}},
\bibinfo{journal}{Phys. Lett.} \textbf{\bibinfo{volume}{B67}},
\bibinfo{pages}{110} (\bibinfo{year}{1977}), \\
\bibinfo{author}{\bibfnamefont{M.}~\bibnamefont{Gell-Mann}},
\bibinfo{author}{\bibfnamefont{P.}~\bibnamefont{Ramond}}
\bibnamefont{and}
\bibinfo{author}{\bibfnamefont{R.}~\bibnamefont{Slansky}},
\bibinfo{book}{Supergravity, 
	  eds. P. van Niewenhuizen and D.Z. Freedman},
\bibinfo{publisher}{North-Holland},
(\bibinfo{year}{1979}), \\
\bibinfo{author}{\bibfnamefont{T.}~\bibnamefont{Yanagida}},
\bibinfo{proceedings}{in Proceedings of the Workshop on the
Baryon Number of the Universe and Unified Theories, Tsukuba, Japan, 
O. Sawada and A. Sugamoto (KEK), 13-14 Feb}, 
(\bibinfo{year}{1979}),  \\
\bibinfo{author}{\bibfnamefont{R.~N.}~\bibnamefont{Mohapatra}}
\bibnamefont{and}
\bibinfo{author}{\bibfnamefont{G.}~\bibnamefont{Senjanov\'{i}c}}, 
\bibinfo{journal}{Phys. Rev. Lett.} \textbf{\bibinfo{volume}{44}},
\bibinfo{pages}{912} (\bibinfo{year}{1980}).

\bibitem{Fukuda:2001}
\bibinfo{author}{\bibfnamefont{S.}~\bibnamefont{Fukuda}},
\bibnamefont{et al. (Super-Kamiokande)}
\bibinfo{title}{Constraints on Neutrino Oscillations Using 1258 Days 
                of Super-Kamiokande Solar Neutrino Data},
\bibinfo{journal}{Phys. Rev. Lett.} \textbf{\bibinfo{volume}{86}},
\bibinfo{pages}{5656}, \bibinfo{note}{hep-ex/0103033},
(\bibinfo{year}{2001}),
\bibinfo{author}{\bibfnamefont{Q.R.}~\bibnamefont{Ahmad}},
\bibnamefont{et al. (SNO)}
\bibinfo{title}{Direct Evidence for Neutrino Flavor Transformation from 
                Neutral-Current Interactions in the Sudbury Neutrino 
		Observatory},
\bibinfo{journal}{Phys. Rev. Lett.} \textbf{\bibinfo{volume}{89}},
\bibinfo{pages}{011301}, \bibinfo{note}{nucl-ex/0204008},
(\bibinfo{year}{2002}),
\bibinfo{author}{\bibfnamefont{Q.R.}~\bibnamefont{Ahmad}},
\bibnamefont{et al. (SNO)}
\bibinfo{title}{Measurement of Day and Night Neutrino Energy Spectra at 
                SNO and Constraints on Neutrino Mixing Parameters},
\bibinfo{journal}{Phys. Rev. Lett.} \textbf{\bibinfo{volume}{89}},
\bibinfo{pages}{011302}, \bibinfo{note}{nucl-ex/0204009},
(\bibinfo{year}{2002}).

\bibitem{Mohapatra:1986}
\bibinfo{author}{\bibfnamefont{R.~N.}~\bibnamefont{Mohapatra}},
\bibinfo{title}{New contributions to neutrinoless double-beta decay 
	  in supersymmetric theories},
\bibinfo{journal}{Phys. Rev.} \textbf{\bibinfo{volume}{D34}},
\bibinfo{pages}{3457} (\bibinfo{year}{1986}).

\bibitem{Kibble:1980}
\bibinfo{author}{\bibfnamefont{T.}~\bibnamefont{Kibble}},
\bibinfo{title}{Some implications of a cosmological phase transition},
\bibinfo{journal}{Phys. Rept.} \textbf{\bibinfo{volume}{67}},
\bibinfo{pages}{183} (\bibinfo{year}{1980}).

\bibitem{Kibble:1995}
\bibinfo{author}{\bibfnamefont{M.}~\bibnamefont{Hindmarsh}}
\bibnamefont{and}
\bibinfo{author}{\bibfnamefont{T.}~\bibnamefont{Kibble}},
\bibinfo{title}{Cosmic strings},
\bibinfo{journal}{Rept. Prog. Phys.} \textbf{\bibinfo{volume}{58}},
\bibinfo{pages}{477}, \bibinfo{note}{hep-ph/9411342},
(\bibinfo{year}{1995}).


\bibitem{Kibble:1982}
\bibinfo{author}{\bibfnamefont{T.W.B.}~\bibnamefont{Kibble}},
\bibinfo{author}{\bibfnamefont{G.}~\bibnamefont{Lazarides}} 
\bibnamefont{and}
\bibinfo{author}{\bibfnamefont{Q.}~\bibnamefont{Shafi}},
\bibinfo{title}{Walls bounded by strings},
\bibinfo{journal}{Phys. Rev.} \textbf{\bibinfo{volume}{D26}},
\bibinfo{pages}{435} (\bibinfo{year}{1982}).


\bibitem[{\citenamefont{Preskill}(1992)}]{Preskill:1993}
\bibinfo{author}{\bibfnamefont{J.}~\bibnamefont{Preskill}}
\bibnamefont{and}
\bibinfo{author}{\bibfnamefont{A.}~\bibnamefont{Vilenkin}},
\bibinfo{title}{Decay of metastable topological defects},
\bibinfo{journal}{Phys. Rev.} \textbf{\bibinfo{volume}{D47}},
\bibinfo{pages}{2324--2342} (\bibinfo{year}{1993}).

%
\bibitem{Chang:1983fu} 
  D.~Chang, R.~N.~Mohapatra and M.~K.~Parida,
  Phys.\ Rev.\ Lett.\  {\bf 52}, 1072 (1984).

\bibitem{Mishra:2009}
\bibinfo{author}{\bibfnamefont{S.}~\bibnamefont{Mishra}},
\bibinfo{author}{\bibfnamefont{U.A.}~\bibnamefont{Yajnik}},
\bibnamefont{and}
\bibinfo{author}{\bibfnamefont{A.}~\bibnamefont{Sarkar}},
\bibinfo{title}{Gauge mediated supersymmetry breaking and the cosmology 
                of the left-right symmetric model},
\bibinfo{journal}{Phys. Rev.} \textbf{\bibinfo{volume}{D79}},
\bibinfo{pages}{065038}, 
(\bibinfo{year}{2009}).

\bibitem{Bajc:2004}
\bibinfo{author}{\bibfnamefont{B.}~\bibnamefont{Bajc}},
\bibinfo{author}{\bibfnamefont{A.}~\bibnamefont{Melfo}},
\bibinfo{author}{\bibfnamefont{G.}~\bibnamefont{Senjanov\'{i}c}}
\bibnamefont{and}
\bibinfo{author}{\bibfnamefont{F.}~\bibnamefont{Vissani}},
\bibinfo{title}{The minimal supersymmetric grand unified theory. 1. 
	  Symmetry breaking and the particle spectrum},
\bibinfo{journal}{Phys. Rev.} \textbf{\bibinfo{volume}{D70}},
\bibinfo{pages}{035007} (\bibinfo{year}{2004}).

\bibitem{Garg:2018}
\bibinfo{author}{\bibfnamefont{I.}~\bibnamefont{Garg}} 
\bibnamefont{and}
\bibinfo{author}{\bibfnamefont{U.}~\bibnamefont{Yajnik}},
\bibinfo{title}{Topological pseudodefects of a supersymmetric $\SO(10)$
		model and cosmology},
\bibinfo{journal}{Phys. Rev.} \textbf{\bibinfo{volume}{D98}},
\bibinfo{pages}{063523}, 
(\bibinfo{year}{2018}).

\bibitem{Rai:1994}
\bibinfo{author}{\bibfnamefont{B.}~\bibnamefont{Rai}}
\bibnamefont{and}
\bibinfo{author}{\bibfnamefont{G.}~\bibnamefont{Senjanov\'{i}c}},
\bibinfo{title}{Gravity and domain wall problem},
\bibinfo{journal}{Phys. Rev.} \textbf{\bibinfo{volume}{D49}},
\bibinfo{pages}{2729}, \bibinfo{note}{hep-ph/9301240},
(\bibinfo{year}{1994}).

\bibitem{Lew:1993}
\bibinfo{author}{\bibfnamefont{H.}~\bibnamefont{Lew}}
\bibnamefont{and}
\bibinfo{author}{\bibfnamefont{A.}~\bibnamefont{Riotto}},
\bibinfo{title}{Baryogenesis, domain walls and the role of gravity},
\bibinfo{journal}{Phys. Lett.} \textbf{\bibinfo{volume}{B309}},
\bibinfo{pages}{258}, \bibinfo{note}{hep-ph/9304203},
(\bibinfo{year}{1993}).

\bibitem{Abel:1995}
\bibinfo{author}{\bibfnamefont{S.}~\bibnamefont{Abel}},
\bibinfo{author}{\bibfnamefont{S.}~\bibnamefont{Sarkar}} 
\bibnamefont{and}
\bibinfo{author}{\bibfnamefont{P.}~\bibnamefont{White}},
\bibinfo{title}{On the cosmological domain wall problem for the 
	  minimally extended supersymmetric standard model},
\bibinfo{journal}{Nucl. Phys.} \textbf{\bibinfo{volume}{B454}},
\bibinfo{pages}{663}, \bibinfo{note}{hep-ph/9506359},
(\bibinfo{year}{1995}).


\bibitem{Mishra:2010}
\bibinfo{author}{\bibfnamefont{S.}~\bibnamefont{Mishra}}
\bibnamefont{and}
\bibinfo{author}{\bibfnamefont{U.A.}~\bibnamefont{Yajnik}},
\bibinfo{title}{Spontaneously broken parity and consistent cosmology 
		with transitory domain walls},
\bibinfo{journal}{Phys. Rev.} \textbf{\bibinfo{volume}{D81}},
\bibinfo{pages}{045010}, 
(\bibinfo{year}{2010}).



\bibitem{Aulakh:1998}
\bibinfo{author}{\bibfnamefont{C.S.}~\bibnamefont{Aulakh}},
\bibinfo{author}{\bibfnamefont{A.}~\bibnamefont{Melfo}} 
\bibnamefont{and}
\bibinfo{author}{\bibfnamefont{G.}~\bibnamefont{Senjanov\'{i}c}},
\bibinfo{title}{Minimal supersymmetric left-right model},
\bibinfo{journal}{Phys. Rev.} \textbf{\bibinfo{volume}{D57}},
\bibinfo{pages}{4174}, \bibinfo{note}{hep-ph/9707256},
(\bibinfo{year}{1998}),
\bibinfo{author}{\bibfnamefont{C.S.}~\bibnamefont{Aulakh}},
\bibinfo{author}{\bibfnamefont{K.}~\bibnamefont{Benakli}} 
\bibnamefont{and}
\bibinfo{author}{\bibfnamefont{G.}~\bibnamefont{Senjanov\'{i}c}},
\bibinfo{title}{Reconciling supersymmetry and left-right symmetry},
\bibinfo{journal}{Phys. Rev. Lett.} \textbf{\bibinfo{volume}{79}},
\bibinfo{pages}{2188}, \bibinfo{note}{hep-ph/9703434},
(\bibinfo{year}{1997}),
\bibinfo{author}{\bibfnamefont{K.}~\bibnamefont{Babu}} 
\bibnamefont{and}
\bibinfo{author}{\bibfnamefont{R.N.}~\bibnamefont{Mohapatra}},
\bibinfo{title}{Minimal supersymmetric left-right model},
\bibinfo{journal}{Phys. Lett.} \textbf{\bibinfo{volume}{B668}},
\bibinfo{pages}{404}, \bibinfo{note}{hep-ph/0807.0481},
(\bibinfo{year}{2008}),


\bibitem{Borah:2011}
\bibinfo{author}{\bibfnamefont{D.}~\bibnamefont{Borah}} 
\bibnamefont{and}
\bibinfo{author}{\bibfnamefont{S.}~\bibnamefont{Mishra}},
\bibinfo{title}{Spontaneous R-parity breaking, left-right symmetry 
	  and consistent cosmology with transitory domain walls},
\bibinfo{journal}{Phys. Rev.} \textbf{\bibinfo{volume}{D84}},
\bibinfo{pages}{055008}, \bibinfo{note}{hep-ph/1105.5006},
(\bibinfo{year}{2011}).

\bibitem{Ferrari:2019}
\bibinfo{author}{\bibfnamefont{S.}~\bibnamefont{Ferrari}},
\bibinfo{author}{\bibfnamefont{T.}~\bibnamefont{Hambye}},
\bibinfo{author}{\bibfnamefont{J.}~\bibnamefont{Heeck}}
\bibnamefont{and}
\bibinfo{author}{\bibfnamefont{M.}~\bibnamefont{Tytgat}}
\bibinfo{title}{$\SO(10)$ paths to dark matter},
\bibinfo{note}{hep-ph/1811.07910},
(\bibinfo{year}{2019}).

\bibitem{GellMann:1961}
\bibinfo{author}{\bibfnamefont{M.}~\bibnamefont{Gell-Mann}},
\bibinfo{title}{The Eightfold Way: A Theory of Strong Interaction 
	  Symmetry},
\bibinfo{note}{Synchrotron Laboratory Report CTSL-20,
	 California Institute of Technology},
(\bibinfo{year}{1961}).

\bibitem{Okubo:1962}
\bibinfo{author}{\bibfnamefont{S.}~\bibnamefont{Okubo}},
\bibinfo{title}{Note on Unitary Symmetry in Strong Interactions},
\bibinfo{journal}{Prog. Theo. Phys.} \textbf{\bibinfo{volume}{27(5)}},
\bibinfo{pages}{949} (\bibinfo{year}{1962}).

\bibitem{Gasser:1980sb} 
  J.~Gasser,
  Annals Phys.\  {\bf 136}, 62 (1981).
  
\bibitem{Jenkins:1991ts} 
  E.~E.~Jenkins,
  Nucl.\ Phys.\ B {\bf 368}, 190 (1992).

\bibitem{Banerjee:1994bk} 
  M.~K.~Banerjee and J.~Milana,
  Phys.\ Rev.\ D {\bf 52}, 6451 (1995)

\bibitem{Aulakh:2005}
\bibinfo{author}{\bibfnamefont{C.S.}~\bibnamefont{Aulakh}}
\bibnamefont{and}
\bibinfo{author}{\bibfnamefont{A.}~\bibnamefont{Girdhar}},
\bibinfo{title}{$\SO(10)$ \`{a} la Pati-Salam},
\bibinfo{journal}{Int. J. Mod. Phys.} \textbf{\bibinfo{volume}{A20(4)}},
\bibinfo{pages}{865--893} (\bibinfo{year}{2005}).

\bibitem{AulakhGirdhar:2005}
\bibinfo{author}{\bibfnamefont{C.S.}~\bibnamefont{Aulakh}}
\bibnamefont{and}
\bibinfo{author}{\bibfnamefont{A.}~\bibnamefont{Girdhar}},
\bibinfo{title}{$\SO(10)$ MSGUT: spectra, couplings and threshold effects},
\bibinfo{journal}{Nucl. Phys.} \textbf{\bibinfo{volume}{B711(1-2)}},
\bibinfo{pages}{275--313} (\bibinfo{year}{2005}).

\bibitem{Aulakh:2004}
\bibinfo{author}{\bibfnamefont{C.S.}~\bibnamefont{Aulakh}},
\bibinfo{author}{\bibfnamefont{B.}~\bibnamefont{Bajc}},
\bibinfo{author}{\bibfnamefont{A.}~\bibnamefont{Melfo}},
\bibinfo{author}{\bibfnamefont{G.}~\bibnamefont{Senjanovi\'{c}}}
\bibnamefont{and}
\bibinfo{author}{\bibfnamefont{F.}~\bibnamefont{Vissani}},
\bibinfo{title}{The minimal supersymmetric grand unified theory},
\bibinfo{journal}{Phys. Lett.} \textbf{\bibinfo{volume}{B588(3-4)}},
\bibinfo{pages}{196--202} (\bibinfo{year}{2004}).

\bibitem[{\citenamefont{Kibble}(1980)}]{Kibble:1980mv}
\bibinfo{author}{\bibfnamefont{T.~W.~B.} \bibnamefont{Kibble}},
  \bibinfo{journal}{Phys. Rept.} \textbf{\bibinfo{volume}{67}},
  \bibinfo{pages}{183} (\bibinfo{year}{1980}).


\bibitem[{\citenamefont{Vilenkin}(1985)}]{Vilenkin:1984ib}
\bibinfo{author}{\bibfnamefont{A.}~\bibnamefont{Vilenkin}},
  \bibinfo{journal}{Phys. Rept.} \textbf{\bibinfo{volume}{121}},
  \bibinfo{pages}{263} (\bibinfo{year}{1985}).


\bibitem[{\citenamefont{Preskill et~al.}(1991)\citenamefont{Preskill, Trivedi,
  Wilczek, and Wise}}]{Preskill:1991kd}
\bibinfo{author}{\bibfnamefont{J.}~\bibnamefont{Preskill}},
  \bibinfo{author}{\bibfnamefont{S.~P.} \bibnamefont{Trivedi}},
  \bibinfo{author}{\bibfnamefont{F.}~\bibnamefont{Wilczek}}, \bibnamefont{and}
  \bibinfo{author}{\bibfnamefont{M.~B.} \bibnamefont{Wise}},
  \bibinfo{journal}{Nucl. Phys.} \textbf{\bibinfo{volume}{B363}},
  \bibinfo{pages}{207} (\bibinfo{year}{1991}).


\bibitem[{\citenamefont{Matsuda}(2000)}]{Matsuda:2000mb}
\bibinfo{author}{\bibfnamefont{T.}~\bibnamefont{Matsuda}},
  \bibinfo{journal}{Phys. Lett.} \textbf{\bibinfo{volume}{B486}},
  \bibinfo{pages}{300} (\bibinfo{year}{2000}), \bibinfo{note}{hep-ph/0002194}.

\bibitem[{\citenamefont{Kawasaki and Takahashi}(2005)}]{Kawasaki:2004rx}
\bibinfo{author}{\bibfnamefont{M.}~\bibnamefont{Kawasaki}} \bibnamefont{and}
  \bibinfo{author}{\bibfnamefont{F.}~\bibnamefont{Takahashi}},
  \bibinfo{journal}{Phys. Lett.} \textbf{\bibinfo{volume}{B618}},
  \bibinfo{pages}{1} (\bibinfo{year}{2005}), \bibinfo{note}{hep-ph/0410158}.

\bibitem[{\citenamefont{Sarkar and Yajnik}(2007)}]{Sarkar:2007ic}
\bibinfo{author}{\bibfnamefont{A.}~\bibnamefont{Sarkar}} \bibnamefont{and}
  \bibinfo{author}{\bibfnamefont{U.~A.} \bibnamefont{Yajnik}},
  \bibinfo{journal}{Phys. Rev.} \textbf{\bibinfo{volume}{D76}},
  \bibinfo{pages}{025001} (\bibinfo{year}{2007}),
  \bibinfo{note}{hep-ph/0703142}.
  
  
\bibitem[{\citenamefont{Coughlan et~al.}(1983)\citenamefont{Coughlan, Fischler,
  Kolb, Raby, and Ross}}]{Coughlan:1983ci}
\bibinfo{author}{\bibfnamefont{G.~D.} \bibnamefont{Coughlan}},
  \bibinfo{author}{\bibfnamefont{W.}~\bibnamefont{Fischler}},
  \bibinfo{author}{\bibfnamefont{E.~W.} \bibnamefont{Kolb}},
  \bibinfo{author}{\bibfnamefont{S.}~\bibnamefont{Raby}}, \bibnamefont{and}
  \bibinfo{author}{\bibfnamefont{G.~G.} \bibnamefont{Ross}},
  \bibinfo{journal}{Phys. Lett.} \textbf{\bibinfo{volume}{B131}},
  \bibinfo{pages}{59} (\bibinfo{year}{1983}).

\bibitem[{\citenamefont{Banks et~al.}(1994)\citenamefont{Banks, Kaplan, and
  Nelson}}]{Banks:1993en}
\bibinfo{author}{\bibfnamefont{T.}~\bibnamefont{Banks}},
  \bibinfo{author}{\bibfnamefont{D.~B.} \bibnamefont{Kaplan}},
  \bibnamefont{and} \bibinfo{author}{\bibfnamefont{A.~E.}
  \bibnamefont{Nelson}}, \bibinfo{journal}{Phys. Rev.}
  \textbf{\bibinfo{volume}{D49}}, \bibinfo{pages}{779} (\bibinfo{year}{1994})

\bibitem[{\citenamefont{de~Carlos et~al.}(1993)\citenamefont{de~Carlos, Casas,
  Quevedo, and Roulet}}]{deCarlos:1993jw}
\bibinfo{author}{\bibfnamefont{B.}~\bibnamefont{de~Carlos}},
  \bibinfo{author}{\bibfnamefont{J.~A.} \bibnamefont{Casas}},
  \bibinfo{author}{\bibfnamefont{F.}~\bibnamefont{Quevedo}}, \bibnamefont{and}
  \bibinfo{author}{\bibfnamefont{E.}~\bibnamefont{Roulet}},
  \bibinfo{journal}{Phys. Lett.} \textbf{\bibinfo{volume}{B318}},
  \bibinfo{pages}{447} (\bibinfo{year}{1993})

\bibitem{Kane:2015jia} 
  G.~Kane, K.~Sinha and S.~Watson,
  Int.\ J.\ Mod.\ Phys.\ D {\bf 24}, no. 08, 1530022 (2015)
  
  
\bibitem{Dutta:2014tya}
K.~Dutta and A.~Maharana, 
  {\emph{Phys. Rev.} {\bfseries D91} (2015) 043503}
  
    %
\bibitem{Cicoli:2016olq} 
  M.~Cicoli, K.~Dutta, A.~Maharana and F.~Quevedo,
  JCAP {\bf 1608}, no. 08, 006 (2016)
  
  
\bibitem{Bhattacharya:2017pws} 
  S.~Bhattacharya, K.~Dutta, M.~R.~Gangopadhyay and A.~Maharana,
  Phys.\ Rev.\ D {\bf 97}, no. 12, 123533 (2018)

\bibitem{Bhattacharya:2017ysa} 
  S.~Bhattacharya, K.~Dutta and A.~Maharana,
  Phys.\ Rev.\ D {\bf 96}, no. 8, 083522 (2017)
  Addendum: [Phys.\ Rev.\ D {\bf 96}, no. 10, 109901 (2017)]
  
\bibitem{Goswami:xyz} R. Goswami and U. A. Yajnik, To appear on ArXiv.

\bibitem[{\citenamefont{Lyth and Stewart}(1995)}]{Lyth:1995hj}
\bibinfo{author}{\bibfnamefont{D.~H.} \bibnamefont{Lyth}} \bibnamefont{and}
  \bibinfo{author}{\bibfnamefont{E.~D.} \bibnamefont{Stewart}},
  \bibinfo{journal}{Phys. Rev. Lett.} \textbf{\bibinfo{volume}{75}},
  \bibinfo{pages}{201} (\bibinfo{year}{1995}), \bibinfo{note}{hep-ph/9502417}.

\bibitem[{\citenamefont{Lyth and Stewart}(1996)}]{Lyth:1995ka}
\bibinfo{author}{\bibfnamefont{D.~H.} \bibnamefont{Lyth}} \bibnamefont{and}
  \bibinfo{author}{\bibfnamefont{E.~D.} \bibnamefont{Stewart}},
  \bibinfo{journal}{Phys. Rev.} \textbf{\bibinfo{volume}{D53}},
  \bibinfo{pages}{1784} (\bibinfo{year}{1996}), \bibinfo{note}{hep-ph/9510204}.

\bibitem{Hong:2015oqa} 
  S.~E.~Hong, H.~J.~Lee, Y.~J.~Lee, E.~D.~Stewart and H.~Zoe,
  JCAP {\bf 1506}, 002 (2015)
  
\bibitem{Hayakawa:2015fga} 
  T.~Hayakawa, M.~Kawasaki and M.~Yamada,
  Phys.\ Rev.\ D {\bf 93}, no. 6, 063529 (2016)
  
\bibitem{Cho:2017zkj} 
  K.~Cho, S.~E.~Hong, E.~D.~Stewart and H.~Zoe,
  JCAP {\bf 1708}, no. 08, 002 (2017)

 
  

\end{thebibliography}
\end{document}